\documentclass[runningheads]{cl2emult}

\usepackage{makeidx}  
\usepackage{graphicx} 
\usepackage{subeqnar} 
\usepackage{multicol} 
\usepackage{cropmark} 
\usepackage{eso}      
\makeindex            

\begin{document}

\title*{Massive Variability Searches:\protect\newline The Past, 
Present and Future}
\titlerunning{Massive Variability Searches}
\author{Bohdan Paczy\'nski}
\authorrunning{Bohdan Paczy\'nski}
\institute{Princeton University, Princeton NJ 08544, USA}

\maketitle              

\begin{abstract}
Many decades ago a search for variable stars was one of the main areas
of astrophysical research.  Such searches, conducted with CCD detectors
rather than with photographic plates, became a by-product of several
projects seeking gravitational microlensing events towards the
Magellanic Clouds and/or the Galactic Bulge: EROS, MACHO, and OGLE.
These searches demonstrated that is is possible and practical to process
in near real time photometry of tens of millions of stars every night, 
and to discover hundreds of thousands of variable stars.  A limited 
subset of new variable star catalogs was published, but no comprehensive 
database of all photometric results became public domain so far.  In the 
last few years a much broader, but shallower searches have been undertaken,
and many other are at various stages of implementation or planning.  There
is a need to develop a system that would allow all these data to be
processed and to be posted on the Internet in real time.  Full information
related to variability of point sources is made of a relatively few data
types, hence it may be relatively easy to handle.  Yet, it may be diverse
enough to be interesting to a large number of users, professional as well
as amateur, making it possible to do real time virtual observing, as well
as data mining.
\end{abstract}

The searches for gravitational microlensing events were undertaken almost
a decade ago, and lead to the first detection of `candidate events' in
1993 \cite{E1}, \cite{M1}, \cite{O1}.  The primary reason for the 
excitement was the hope that this may be a way to learn about dark matter,
if it were composed of massive compact objects \cite{P1}.  This hope was
met with scepticism about technical feasibility of those projects:
there had been no precedent for processing so much data in near real time
in optical astronomy.  Seven years later the technical scepticism is gone,
but there is almost a consensus that the majority of microlensing events
are caused by ordinary stars, some brown dwarfs, but the lensing objects do 
not contribute significantly to the dark matter \cite{E2}.

The main and lasting impact of microlensing searches has been
the development of hardware and software capable of handling tens of millions
of photometric measurements every clear night, archiving the data,
real time recognition of the very rare microlensing events \cite{O2},
\cite{M2}, and the distribution of alerts to all interested observers,
as well as posting them on the WWW: \hfill
\break
\centerline{http://www.astrouw.edu.pl/\~~ftp/ogle/ogle2/ews/ews.html}
\centerline{http://darkstar.astro.washington.edu/}
\noindent
These two alert systems are not active now, as MACHO ended its project 
on December 31, 1999, and OGLE is shut down for a few months for a major
upgrade.  OGLE alerts, known as EWS (Early Warning System) will resume
in late 2001.

The microlensing searches monitored tens of millions of stars over several
years, typically once every clear night, and therefore hundreds of data
points per star.  This amounts to $ \sim 10^{10} $ photometric measurements.
More than $ 10^5 $ of the stars were found to be variable.  While catalogs
of thousands of variables were published, and data is available on the WWW,
e.g. \hfill
\break
\centerline{ http://bulge.princeton.edu/\~~ogle/ }
\centerline{ http://www.astrouw.edu.pl/\~~ftp/ogle/index.html }
\noindent
no comprehensive archive of these
results is available in public domain.  OGLE collaboration is making an
effort to create such an archive, motivated to a large extent by the upgrade
from OGLE-II ($ 2k \times 2k $ CCD camera with 0.4'' pixels) to OGLE-III 
($ 8k \times 8k$ CCD camera with 0.25'' pixels).  It is clear that as soon 
as the much larger flood of higher quality OGLE-III data begins to flow,
nobody will have time to look back at the OGLE-II data, which will contain
a lot of unexplored science.  Hence the need to make the `old' data public
domain as soon as possible.  A sample of the format to be expected for the
archive is provided by the DC/AC presentation of one of the Bulge fields 
\cite{W1}.  The DC part is a FITS file containing a high S/N reference
image of the field with $ 2k \times 8k $ pixels, obtained by co-adding
20 of the best images obtained during three observing seasons: 1997, 1998,
and 1999.  The AC part is a catalog of photometric measurements for
$ \sim 4,600 $ variables for all $ \sim 250 $ difference images obtained
by subtracting the reference image from individual CCD frames.  The virtue
of this decomposition is a reduction of the volume of data from $ \sim 8 $ 
GB for raw CCD images to $ \sim 110 $ MB for the DC/AC representation,
a compression by a factor $ \sim 80 $, making it practical to put the 
DC/AC data on line at: \hfill
\break
\centerline{ http://astro.princeton.edu/$\sim$wozniak/dia }
\noindent
We plan to archive in this way all 4 seasons of OGLE-II results,
which corresponds to $ \sim 1 $ TB of raw CCD images, to be converted
to $ \sim 12 $ GB of the DC/AC data.  Naturally, there is some
information loss in this compression process, and it remains to be seen
if the proposed archiving scheme turns out to be practical.

The microlensing projects and other massive variability searches
left a misleading impression that
it takes a team of dozens of collaborators to implement and to operate them.
Indeed, there are plenty of examples: EROS,
MACHO, MOA, ROTSE, etc.  However, there are also counter examples, a small
OGLE team, and single person DUO and ASAS.  Links to the Web sites of most
of these projects, and many other small, mostly robotic instruments
can be found at: \hfill
\break
\centerline{ http://www.astro.princeton.edu/faculty/bp.html }
\centerline{ http://alpha.uni-sw.gwdg.de/\~~hessman/MONET/ }

As far as I can see there is nothing intrinsic to massive variability
searches that requires a big team.  The size is determined by the choice 
of the PI and/or the circumstances under which a project develops.  The 
structure of the undertaking, is the project `open' or `closed', is it 
focused or broad, all vary from case to case.  I like an approach with
a large number of small teams, open projects, with a minimum of 
central coordination but with full information exchange.  The tasks are
very diverse: instrument development (hardware and software),
observations, data processing, archiving, analysis.  All these can be done
within a single, huge team, or different tasks may be carried out by many
independent small teams and individuals.  It is far too early to tell which
approach: top down and autocratic, or bottom up and democratic, will be more
productive and sustainable over a long time interval.  I certainly prefer
the latter.

It is not generally appreciated how under-explored is the sky of bright
objects \cite{P2}, \cite{P3}.
There are more than 10 big new telescopes, operating or under 
construction, with aperture in the range 6.5 - 10 meters, built to
explore the faintest possible objects, with cosmology being the the primary
scientific topic.  Most of them have very small fields of view, but some
have wide fields up to half a degree across.  It will take over $ 10^5 $
of such `wide field' images to cover the whole sky with a single filter.
While cosmology is one of the most exciting areas of science, and it is 
natural to investigate the ever fainter objects at the ever higher angular and
spectral resolution, it is a mistake to ignore the bright sky.  It is commonly
assumed that there is nothing of importance to be discovered at the
bright end.  Yet, in 1999 two out of ten most important and most spectacular
astronomical discoveries were made with 10 cm instruments: an optical
flash from GRB 990123 from the redshift of 1.6 \cite{R1}, and the 
first detection of a planetary transit in front of its star \cite{p2}
(the transit has also been discovered by a larger robotic telescope 
\cite{p1}).

Making a major discovery, like the two that I have just mentioned, is
like winning a lottery.  However, less spectacular, but important and 
guaranteed discoveries of thousands of new variable stars can also be
made with a 10 cm instrument, as demonstrated by ASAS \cite{A1} and ROTSE
\cite{R2}.  All these were brighter than 13 mag, and 90\% of them were new.
A simple extrapolation indicates that there are $ \sim 10^5 $ bright
variables waiting to be discovered using 10 cm instruments.  It demonstrates
how incomplete is our knowledge of the bright sky.  I think this
ignorance is inexcusable and embarrassing to the astronomical community.

It is important to have a complete census of all variable objects 
which are readily accessible to small telescopes.  Complete catalogs
of variable stars are useful for the studies of stellar evolution as well
as galactic structure.  They are needed to recognize
any new type of variability, like an optical flash or a planetary transit,
using small instruments alone, without the need to be alerted by the
results obtained with large telescopes.  Note, that the optical flash
from GRB 990123 would not have been recognized by the ROTSE team if not
for the small error box provided by the BeppoSAX \cite{SAX}.  The star
selected for the planetary transit search was preselected on the basis
of spectroscopic data obtained with a full size telescope \cite{p3}.

It is technologically possible to search the whole sky for bright optical
flashes with fully robotic instruments.  There are many reports of such 
flashes \cite{f1}, \cite{f2}, \cite{f3}, but they are impossible to verify
years after the event has been observed.  The hardware needed for the task
is readily available, as demonstrated by ASAS, LOTIS, ROTSE, TAROT, and many
others groups.  What is missing is software capable of
real time data processing and automatic recognition of anything unusual
happening in the sky, and alerting other robotic instruments to carry out
the verification and a more detailed study.  A similar system worked
brilliantly with GRB 990123, with the alert system developed for 
the BATSE on Compton GRO, and followed up robotically by ROTSE \cite{R1}.
Near real time alert systems were developed
for supernovae and microlensing events, but they do not exist for rapid
variability, flashes which may last only minutes or even seconds.  No doubt
it is difficult to develop the necessary software, but the most important
problem is the lack of will to develop it.  The problem seems to be
cultural, rather than scientific or technological.

In the past few decades there were many excellent space projects designed
to monitor all sky in gamma-ray and X-ray domains, and a large number of
new types of events and variables were discovered, including gamma-ray
bursts, X-ray bursts, X-ray pulsars, X-ray novae, etc.  There were also many
all sky (or nearly all sky) imaging projects in the radio, infrared and
optical domains.  However, the only all sky optical variability search
was done serendipitously by the Hipparcos project \cite{H1}, limited to
stars brighter than $ \sim 10 $ mag.  Obviously, the all sky search for
objects which vary in the optical
domain is made difficult by the presence of millions of stars which do
not vary, as well as thousands of `boring' or `ordinary' variables.  But the
hardware that is required is so much less expensive than hardware for
X-ray and gamma-ray instruments, that it should compensate for the annoying
background and foreground of not so interesting sources.  Of course, for
many of us those are also interesting.  The following is
a brief description of the current status and the plan for expansion of
the All Sky Automated Survey (ASAS), as developed by Dr. G. Pojma\'nski
of the Warsaw University Observatory.  Many details may be found at: \hfill
\break
\centerline{ http://archive.princeton.edu/\~~asas/ }
\centerline{ http://www.astrouw.edu.pl/\~~gp/asas/asas.html }

ASAS begun in 1997, with a single instrument placed at the Las Campanas
Observatory in Chile.  The optics was a f/1.8 lens with $ f = 135 $ mm, and
the detector was the Meade/Pictor 416 CCD camera with $ 512 \times 768 $
pixels.   The camera and the equatorial mount were under full computer
control.  The instrument accumulated data from 50 fields, each $ 2 \times 3 $
degrees, for the total area $ 300 $ square degrees, i.e. about 0.7\% of
the sky.  During 2.5 years of operation several hundred images were obtained
for each field, and a total of $ \sim 140,000 $ stars were recorded, down to
$ I \approx 13 $ mag.  Approximately 3,900 of these varied, most of them
on a long time scale, presumably being Mira-type or 
semi-regular red giants.  More than 90\% of variables were new discoveries,
demonstrating how incomplete is our knowledge of the bright sky.

In July 2000 the ASAS was upgraded with three new instruments, each 
equipped with an Apogee AP10 CCD camera with $ 2k \times 2k $ pixels.
Two cameras are attached to Minolta lenses with $ f = 200 $ mm, f/2.8,
with standard V-band and I-band filters, respectively.  Each of these
instruments images all sky accessible from Las Campanas once every two
nights.  A third instrument has a mirror f/3 aperture with the diameter of
25 cm, a field of view 2 degrees across, and an I-band filter.  It images
a small fraction of the sky only.  At this time the observations are archived,
and the fully automatic data pipeline is under development.  Preliminary
analysis indicates that photometric accuracy is satisfactory, but two of
the cameras have unacceptably large dark current.

While ASAS will search all southern sky for variability on a $ \sim 1 $ day
time scale it would be interesting to monitor the sky more frequently, say
every hour, or even every minute, down to whatever magnitude is
possible with the available resources \cite{N1}.  The first few CONCAMs were
developed by Dr. R. Nemiroff with this goal in mind.  As it is often the
case software is more challenging than hardware, hence only limited search
for variability has been conducted so far \cite{N2}, but the instruments have
already found their application at several observatories as weather monitoring
devices.  There is every reason to expand the capability of CONCAM-like
instruments, but it is clear that software development will be the most
challenging part of the undertaking.

I expect, or rather I hope, that ASAS will gradually expand the frequency
and the depth of its sky coverage, with gradual development of real
time software, full alert system, and a long term data archive, suitable
for data mining.  Note, that variability implies that it may be sufficient
to restrict the database to point sources, which means very simple and
not diverse data types.  This may be easier to handle than a much broader
concept of a Virtual Observatory discussed at this meeting.  However, the
diversity of variability types may be rich enough to attract users.
It remains to be seen if there will be a market for virtual observing,
i.e. will the database be used even if virtual observers are not provided
with financial incentives to access and to analyze the data.

I envision the progress of ASAS as a step by step development of real time
software. First, a catalog of all
stars detectable on most CCD frames will be compiled and variable stars
will be cataloged and classified.  Next, any new object will be recognized
by its absence in the archive.  It is not known how many new objects will
be detected every night, and what fraction of those will be instrumental
artifacts, atmospheric events, or man made events (satellites, airplanes,
balloons, etc).  Presumably there
will be a usual learning curve in recognizing and classifying new objects.
Once non-interesting cases are identified and rejected, the potentially
interesting cases will be subject to a robotic verification.  Once their
reality is established electronic alerts will be sent by e-mail and posted
on the WWW.  It is not clear at this time who will develop the intelligent
software capable of real time recognition of new and/or interesting events:
with many kinds of variability in light intensity or position.  ASAS is an open
project, and all competent and interested collaborators are most welcome.
It is useful to note, that it is much easier to follow up and analyze a 
bright new object, than it is to do the same for a 25 mag source.  Yet,
we do not know at which flux levels more unexpected and more interesting
events will be discovered.  The domain of fast optical transients remains
unexplored.

This project was supported in part by the NSF grants AST-9820314 and
AST-9819787.

\end{document}